# Security and Privacy Concerns in Cloud-based Scientific and Business Workflows: A Systematic Review

Nafiseh Soveizi*, Fatih Turkmen*, Dimka Karastoyanova*

*University of Groningen, Groningen, The Netherlands

**Abstract:** Today, the number of data-intensive and compute-intensive applications like business and scientific workflows has dramatically increased, which made cloud computing more popular in the matter of delivering a large amount of computing resources on demand. On the other hand, security is a critical issue affecting the wide adoption of cloud technologies, especially for workflows that are mostly dealing with sensitive data and tasks. In this paper, we carry out a review of the state-of-the-art on how security and privacy concerns in scientific and business workflows in cloud environments are being addressed and identify the limitations and gaps in the current body of knowledge in this area. In this extensive literature review, we first present a classification of the state-of-the-art security solutions organized according to the phases of the workflow life cycle they target. Based on our findings, we provide a detailed review and classification of the most relevant available literature focusing on the execution, monitoring, and adaptation phases of workflows. Finally, we present a list of open research issues related to the security of the cloud-based workflows and discuss them.

## 1. Introduction

Workflows are commonly used application models that consist of a series of computational tasks logically connected by data- and control-flow dependencies [1]. There are two main types of workflows: The first type – *scientific workflows*, typically involve a large amount of data processing, analysis, and computing, requiring high computing and storage capacities. The second type of workflow is *business workflows*, applied predominantly in (business) information systems. Unlike scientific workflows, individual activities/tasks in business workflows usually require lower computing power and fewer storage resources, although the number of concurrently running workflow instances is typically large and the communication time between tasks needs to be kept as short as possible [2]. Both academia and industry have been active in setting the foundations of workflow management (as a subfield of Business Process Management) as a discipline in the last several decades by developing the workflow technology that allows for modelling, executing workflows, and thus automating the enterprise processes.

Cloud computing [3] plays a key role in workflow management since it can deliver a large amount of computing resources on-demand [2] for running data-intensive and compute-intensive applications. It also comes with the promise to reduce the running costs and maximize the revenues while maintaining or even improving the Quality of Service (QoS). Making use of cloud computing by Workflow Management Systems (WfMSs) can further increase the productivity of the system. Seen from the point of view of cloud computing users, cloud workflows [4][5] provide an abstract definition of complex applications, flexible configuration, and automated scalable operation, and also improve the QoS.

From the perspective of the providers of cloud computing services, cloud workflows enable the automatic scheduling of tasks (the process of mapping tasks to cloud resources within the required QoS) and management of resources [6].

Despite all the above-mentioned advantages of cloud-based workflows, cloud security is a major area of concern [7][8] that is restricting their use for certain applications, especially for the workflows dealing with sensitive data and tasks. In fact, when a workflow or part of it is outsourced to the cloud, the WfMS loses control over tasks that can lead to increased security risks and make them vulnerable to malicious attacks. These security challenges mostly stem from the shared nature of cloud infrastructure in which computing/storage resources are shared with other users, and sensitive data are transferred among cloud components such as Data Centers (DCs) over possibly untrusted network channels. In addition, the cloud is honest-but-curious in the sense that the cloud service provider may faithfully follow the established protocols but at the same time, it may be curious to deduce valuable information about the users' data and the workflow logic. Since the deduced information may be leaked or even sold to third parties by the malicious cloud providers [9], some users are reluctant to use the cloud (deployment model).

In search for a solution to these concerns, there have been various studies on the topic of security properties of processes and workflow management systems from different perspectives. To the best of our knowledge, there is no up-to-date overview of the state of the art on the topic. To compensate for this gap in existing research, in this work we use the methodologies for performing a literature mapping study and a systematic literature review to achieve two goals:

1) Systematic review of the state of the art in addressing and maintaining security properties of cloud-based workflows throughout their complete life cycle, including a special focus on an additional life cycle phase accounting for runtime adaptation.

2) Identification of the gaps in the state of the art and subsequently the needs for future research.

Our study shows that most of the solutions for ensuring the security properties of cloud-based workflows focus on the modelling aspects of workflows, mainly providing modelling concepts to express the security properties of workflows. Solutions towards enforcing these properties are rare, narrow in scope, and in some cases implementation-specific. One significant finding is that there is only very scarce research reported on the runtime adaptation of cloud-based business or scientific workflows that is triggered upon and carried out as a reaction to security violations.

The rest of the paper is organized as follows: Section 2 introduces the basic concepts. In Section 0, we discuss the existing literature about the security and privacy of cloud-based workflows. Section 4 presents the review process and data collection. In Section 5, the main results of our study are presented. Section 6 illustrates the open issues and the challenges that still need to be addressed in this topic. Finally, Section 7 concludes this paper.

## 2. Background: Scientific and Business Workflows, and the Similarities/Differences between them

The Business Process Management (BPM) (that covers also the workflow management technology) and scientific workflows are established research fields and up until recently have been regarded as separate fields of research. In the last decade, there have been several attempts to apply approaches from the BPM field in the field of scientific workflows both in terms of modelling approaches and principles, as well as in terms of using workflow management environments to run scientific workflows for different application fields [10] [11]. At the same time, these works were based on the observations that there are both similarities and differences between these seemingly disparate fields. In this section, we highlight these differences and provide the necessary background information on the topic.

Based on the available literature, we can summarize the definitions of both terms as follows:

A *scientific workflow* describes a series of computations that enable the analysis of data in a structured and distributed manner. It orchestrates and automates scientific applications in a way that reduces the complexity of managing scientific experiments [12].

A *business workflow* is the automation of a business process, in whole or in part, during which documents, information, or tasks are passed from one participant to another for action, according to a set of procedural rules [13].

The life cycles of workflows that both fields follow have a different focus as depicted in Figure 1. The two life cycles clearly show that the two fields view the workflows from different perspectives: business workflows are viewed as a software artifact that can be used by several user roles with focus on different aspects of the management of the artifacts, whereas the scientific workflows revolve around one user role, namely the scientist, who is dealing with both management of the software artifacts and their use. Note that we will mostly refer to the business process life cycle in our study, as it is the more detailed one and hence provides a more detailed basis for the comparison of the existing works. Despite seemingly different focus of the two fields, the literature shows [12][14][10][11][15][16] similarities as well, which were the reasons for the recent technological advances mentioned above. Based on these similarities especially in the security concerns, we investigate both types for workflows together in this survey. Table 1 and Table 2 summarize the similarities and the differences between scientific and business workflows respectively.

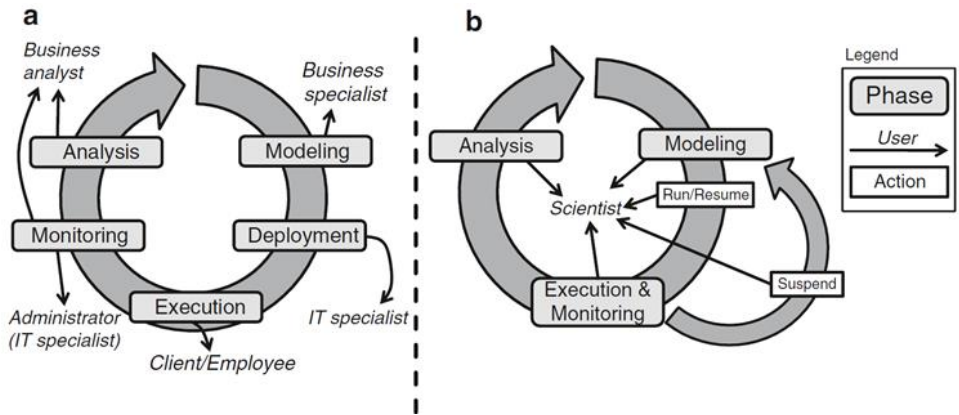

Figure 1: a) Business Workflow Life Cycle b) Scientific Workflow Life Cycle [10]

Table 1: The Similarities between Scientific and Business Workflows

| Criterion | Business and Scientific Workflows |
| --- | --- |
| **Security** | Both types of workflows need to satisfy the fundamental security principles of CIANA (Confidentiality, Integrity, Availability, Non-Repudiation, and Authenticity) [17] during the whole workflow life cycle. |
| **Robustness** | The requirements like the ability to be error-resistant and recoverable are similar for both types of workflows. |
| **Scalability** | Both types of workflows require the ability to scale with the number of users, services, data resources, and involved participants. |

Table 2: The Differences between Business and Scientific Workflows

| Criterion | Business workflows | Scientific workflows |
| --- | --- | --- |
| **Workflow definition and execution** | Business processes typically define control and data flow in a process model using a generic (domain-independent) notation. They are executed multiple times on a generic process execution environment. | Scientific workflows are defined either using a programming language or a domain-specific scientific workflow language or notation. Execution is system-specific too. |
| **Data flow vs. control flow** | Control-flow oriented, focus on tasks/activities and their ordering. | Data-flow oriented, explicit focus on data and its processing. |
| **Life Cycle** | - Explicitly defined life cycle where phases focus on managing processes/workflows<br>- One model and many instances<br>- Different groups of users<br> (Fig 1. a) | - Life cycle phases focus on managing the scientific computation from point of view of the user<br>- No explicit distinction between workflow models and their instances<br>- Scientists are the only user group<br> (Fig 1. b) |
| **Duration** | - Short and long running processes<br>- Number of instances may be huge | - Short and long running computations<br>- Number of instances smaller |
| **Flexibility** (a.k.a. dynamicity) | Usually, workflows are pre-defined during the modeling phase. Mostly academic research results in process evolution and adaptation available. | Need a high degree of flexibility because they are carried out in a trial-and-error manner. |
| **Reproducibility** | Less need for reproducibility | Need reproducibility |
| **Fault Handling** | Processes must be guaranteed to be complete and if any fault occurs, it should be handled. Means are available for fault and exception handling in the existing technologies. | Conduct experiments that may or may not succeed. However, technical faults (e.g. server unavailability, network connection error) that may occur during the execution should be handled. FH and EH on scientific workflow level specific for the domain language, if at all available. |
| **Interaction with participants** | Data can be processed by machines or humans. In most cases, several users are involved. | In most cases, data is processed only by machines, and the scientists just manage and monitor the workflow execution. |

## 3. Related Surveys

Reviews are typically divided into two types: Systematic literature Review (SLR)[18] and Traditional Literature Review (TLR)[1] [19][20]. SLRs usually try to answer well-defined questions by following a specific search strategy. On the other hand, TLRs usually do not mention their search strategy for finding relevant publications, and therefore in TLRs, searches may be ad-hoc and are thus not fully comprehensive.

There are very few literature reviews that can be related to the security and privacy of cloud-based workflows. In Table 3, we provide an overview of the relevant literature reviews.

The role of trust in service workflows has been examined and explored in [21]. The authors have defined trust as a complement to conventional security services (e.g., authentication, authorization). Therefore, the main focus of this paper is trust that can improve security in business workflows where security requirements are locally defined, globally integrated, and distributedly enforced. Based on their findings, workflows need to be more flexible in terms of trust mechanisms to enable an increase in the degree of automation.

The survey [6] that is closest in scope to our survey, provides an initial overview of cloud workflow security. It has mapped the specification of QoS to the workflow life cycle phases as follows: The QoS specification is done in the workflow modeling stage; QoS aware service selection happens in the instantiation stage of the workflow where the appropriate software and hardware services are selected based on the requirements specified in the previous stage; and QoS consistency modeling and QoS violation handling happen in the workflow execution stage. However, based on the publication year (2014) of this paper and also the type of its review, it does not provide a comprehensive overview of the recent developments.

The TLR presented in [22] has surveyed the existing works by defining the factors needed in securing scientific workflows during execution, identifying several domains in which security is essential and sources of security threats. The paper only focuses on the scheduling phase of the scientific workflow.

In [23], the security concerns in resource scheduling have been investigated. The authors identified the different types of security constraints and classified models into three categories: data security, data center security, and infrastructure security. The focus of this paper is only on the scheduling phase. These literature reviews have different goals and/or do not cover all phases of the workflow life cycle. Hence, we can conclude that there is a lack of a comprehensive study of the security and privacy concerns of the cloud-based workflows during the whole workflow life cycle and their effect on the WfMS architecture.

Table 3: The summary of the Related Literature Reviews.

| Paper | Type of review | Focus | Main findings |
|---|---|---|---|
| [21], 2012 | TLR | Trust exhibited in service workflows (trust is considered as a complement to the conventional security services) | Formal definitions of trust need more study to be usable as means for decision making in dynamic distributed environments and as a result, increase the degree of automation. |
| [6], 2014 | TLR | An overview of cloud workflows and security | There must be cloud-specific standards for securing the workflows in the cloud. |
| [22], 2018 | TLR | Security of the scientific workflows during execution | There is a need of developing more models which will consider different parameters such as (execution) environment, CPU configuration settings, for more than one workflow. |
| [23], 2019, covers 2006-2015 | SLR | Security concerns in resource scheduling | The main focus of their reviewed studies is limited to Integrity, Availability, and Security. |

---

[1] Narrative review

## 4. Research Methodology

As mentioned before, up to now and to the best of our knowledge, there is no comprehensive review that can discover and evaluate the security and privacy concerns in cloud-based business or scientific workflows. To overcome this, we use a combination of an SLR [24] and a Systematic Mapping Review (SMR) [25] to identify the current research challenges and also existing gaps that can give an overview of research in the area. Since the articles are not evaluated in such detail in practice according to the SMR protocol, more articles can be considered. For that reason, as a first step, we used SMR to portray the relationship between literature and categories and identify gaps, and show in which topic areas there is a shortage of publications [25]. Subsequently, we use the mapping as a road map for the next steps, namely an SLR, with which we show further details about existing works on the identified research question. Our Review Methodology Structure is presented in Figure 2.

### 4.1. Research Questions

The research questions we define for our research are as follows:

**RQ1**: What is the state of the art in security and privacy in cloud-based business and scientific workflows in each stage of their life cycle?

**RQ2**: Which security issues are addressed in which phase of the life cycle and what mechanisms are employed?

**RQ3**: Based on the research identified, what are the existing research gaps on which further research should focus?

### 4.2. Search Strategy

The search was performed in four scientific databases, namely Scopus, Web of Science, ACM Digital Library, and IEEE Xplore. We also scanned the reference lists included in the papers in order to ensure that this review would be more comprehensive. The search was limited to papers in English published between January 2010 and December 2021.

The initial search string used to find the related papers is as follows:

*(("security" OR "privacy") AND ("scientific workflow" OR " "business workflow"" OR "business process" OR "service Composition" OR "orchestration") AND ("cloud"))*

In order to cover all papers that are related to the same or similar concepts in the literature, especially in the business context, we also used "business process", "service composition" , and "orchestration" in the search string.

### 4.3. Study Selection Criteria and Procedures

This section describes the inclusion/exclusion criteria that set the boundaries for the systematic review and also the procedures for performing the selection.

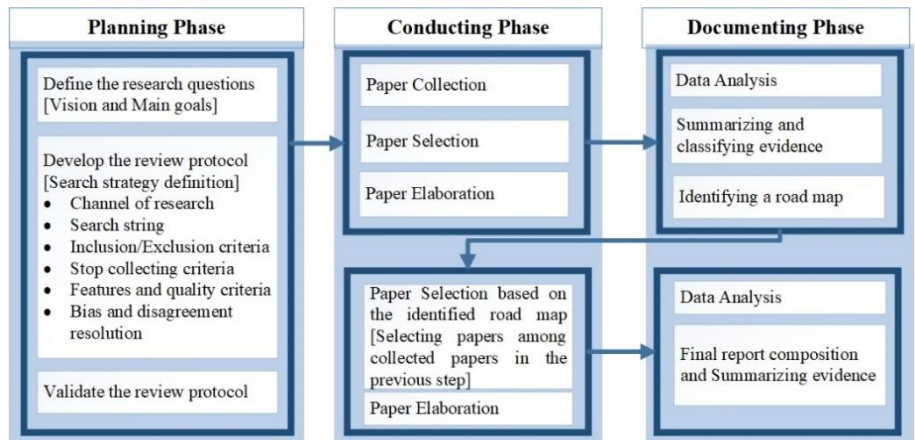

Figure 2: Review Methodology Structure (based on [24])

#### 4.3.1. Inclusion/Exclusion

The inclusion criteria for the selection of papers are:
- Published Conferences, Workshops, Journals, and peer-reviewed papers that address any aspect of security or privacy in cloud-based business or scientific workflows in one or more stages of their life cycle.
- Any previous literature reviews in this area.

The exclusion criteria are:
- Studies with inadequate information that are only available as presentations, abstracts, or are otherwise incomplete.

  Duplicate reports of the same study. When several reports of a study exist in different journals, the most complete version of the study was included in the review.
- Forms of publication that have not been subjected to a formal review process (non-peer reviewed literature or gray literature), including journals such as ACM Software Engineering Notes (unless containing conference proceedings) and technical reports.
- Opinion papers.

#### 4.3.2. Procedures for Selection

After removing the duplicate publications and conference announcements, we reviewed the titles, abstracts, and keywords of the studies, and the approved studies were selected for further analysis based on the inclusion/exclusion criteria. The study design process is depicted in Figure 3.

## 5. Results And Discussion

In this section, we provide a summary of the results of the study and discuss our main findings.

### 5.1. Security Challenges and Solutions

The papers found in the first phase of the review (120 papers) have tried to address the different security objectives of the workflows in the cloud environments. Figure 4 shows the percentage of the objectives covered by the papers. As the chart shows, Confidentiality (data and logic), Integrity (data and task), and Availability (CIA) are the most important security properties considered in the literature.

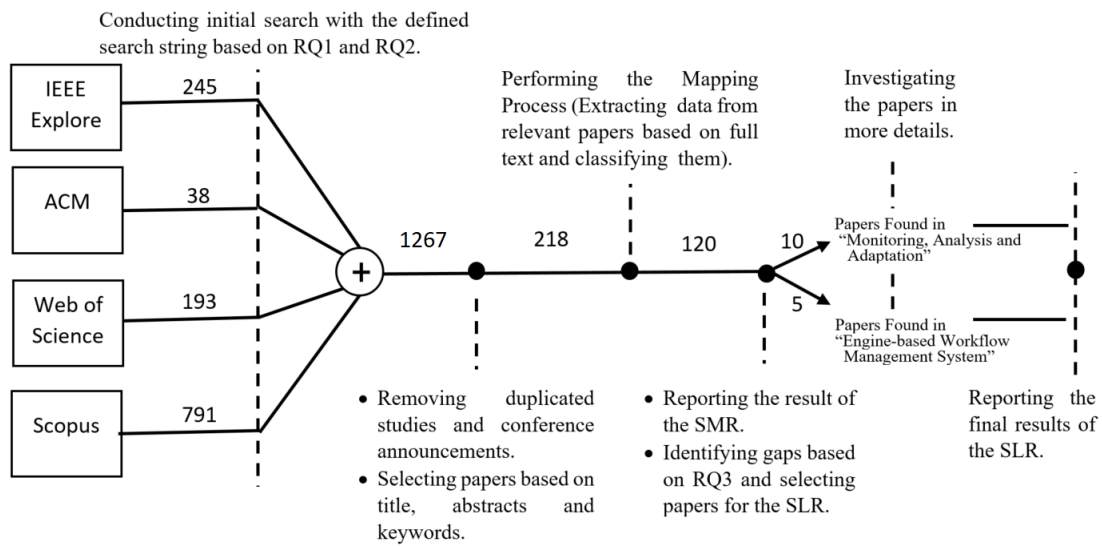

Figure 3: The study design process

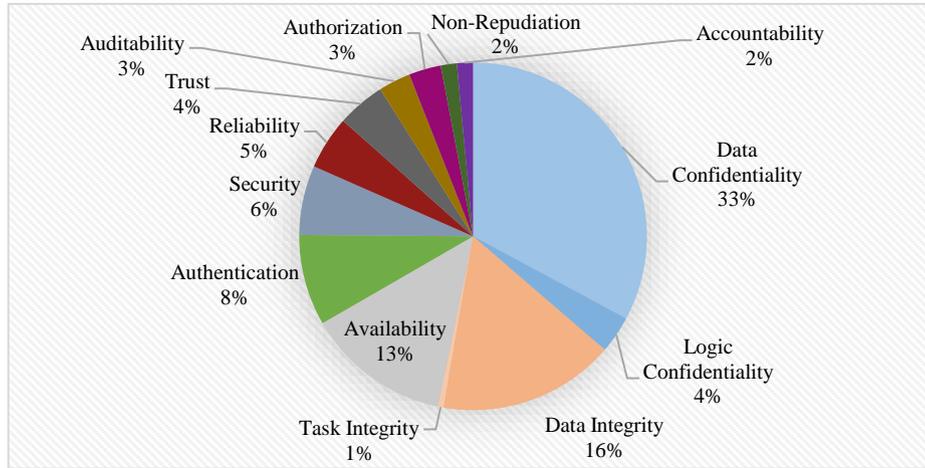

Figure 4: Percentage of Security Objectives covered by the papers.

Some of these research works (109 papers) have tried to provide certain solutions to achieve these objectives. Figure 5 depicts the classification of the selected papers based on their proposed solutions in the context of cloud computing environments.

We have organized the provided solutions in groups as follows:

### 5.1.1. Virtualization and Security Services

Usually cloud providers offer different security services and various levels of isolation guarantees for each service. For example, different availability methods (e.g. protective redundancy models and overload protection), encryption (SEAL, RC4, RC5, and IDEA), integrity services (e.g. hash functions like MD4, MD5, RIPEMD, RIPEMD-128, SHA-1, RIPEMD-160, and Tiger), and authentication mechanisms (e.g. HMAC-MD5, HMAC-SHA-1, and CBC-MAC-AES) are used to meet the availability, confidentiality, integrity, and authentication goals, respectively. These security services can be applied based on the level of security required by the user. Also, in order to provide confidentiality and user privacy in the workflow engine and/or on the client-side before deploying the process into an engine, different obfuscation and diversification techniques are proposed:

a) *Data Obfuscation*: In this method, the confidentiality of the data is achieved by encrypting the data or obfuscating it on the client-side (e.g., splitting data, noise injection, and deleting sensitive data) before sending the data to the cloud;

b) *Diversification*: This method tries to diversify the cloud execution environment continuously and shorten the chance and the time for the attacker to discover the execution environment and its vulnerabilities. In other words, before the attacker acquires the knowledge about the execution environment, it would already be changed to a new one in order to render the acquired knowledge obsolete [26]. Diverse physical resources like servers, hypervisors, operating systems, or the workflow execution environments can be used to confuse the attackers [27];

c) *Logic obfuscation* (BP obfuscation): In this method, the user or broker tries to split the process (splitting the BP model in a choreography of BP (fragments)) so that each cloud has only a partial view of the model.

d) *Information Flow Checking*: This method tries to quantify the information flow to evaluate the intra-service leakage between different inputs and outputs of each service. It also aims to ensure inter-service flow security by evaluating the candidate services in the service chain [28].

### 5.1.2. Administrative Decision

These decisions help users to express and verify the security and privacy requirements or strike a balance between these requirements and other possible user preferences like execution time and cost. They can be divided into the following three categories:

a) *Extending Modeling and Execution Tools*: The papers in this group mostly focused on extending languages and/or

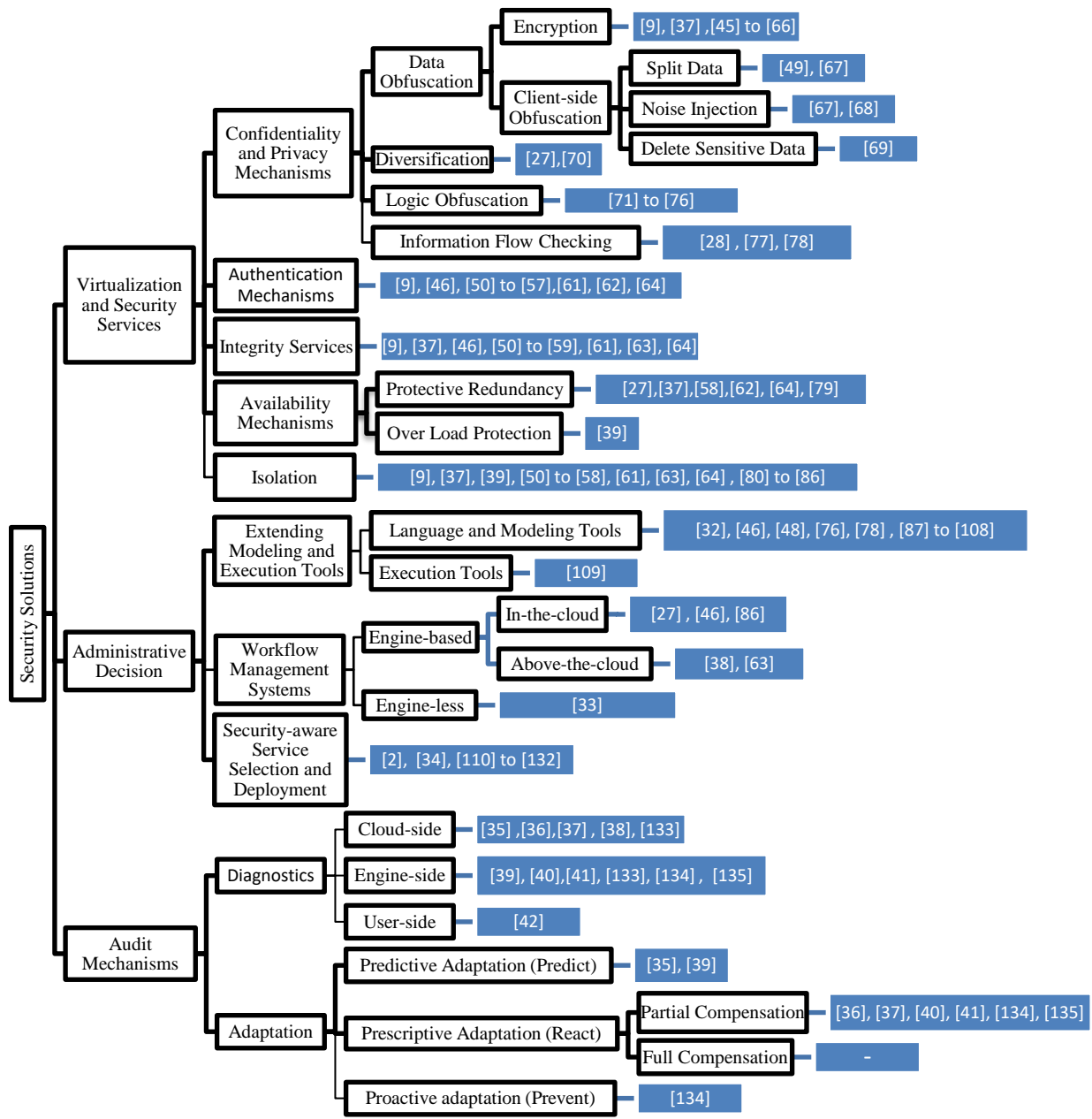

Figure 5: Classification of the Security Solutions introduced by the corresponding

modeling tools to specify the user security and privacy requirements. For instance, some of them try to capture the access control requirements in business and scientific process specifications and then tried to establish mechanisms to enforce these requirements, like supporting the principle of strict least privilege, the delegation of authority, the integrity principle, scalability, efficiency, and revocation [29], [30], [31]. Furthermore, a few papers try to verify the security of the workflows based on the predefined secure service composition (SCO) patterns [32].

b) *Workflow Management System*: These papers addressed the architecture of the WfMS that can handle the security and privacy requirements. Most of the proposed WfMSs are engine-based. It means that they have an engine that is responsible for controlling the execution of the workflows. This engine can be placed in the private cloud or user-end system (i.e., Above-the-cloud) or it can be fully deployed in the public cloud (In-the-cloud) and thus the cloud has full control over the execution and monitoring of the workflows [6]. WfMS can also be centralized (controlling the workflow execution from one location in a single engine) or distributed (multiple workflow engines that provide the ability to cope with peaks in the system load and distributed environment) [101]. However, to the best of our knowledge, there is no distributed cloud-based WfMS. More details about these engine-based security-aware WfMSs are discussed in Section 5.4. Besides, the paper [101] introduces the engine-less WfMS. The key idea of this kind of engine is to have the workflow process instance be self-protected and not need a workflow engine to secure the data therein. A detailed discussion on the engine-less WfMSs is out of the scope of this paper.

c) *Security-aware Service Selection and Deployment*: These methods try to select services based on the user requirements. They must be able to make a balance between different user demands such as time, cost, and security during the scheduling of the processes. These papers addressed different aspects of security challenges during the scheduling of the workflows like confidentiality, integrity, authentication, availability, reliability, and trust. Most of them considered the level of employed security methods for each deployed VM that can fulfill the defined security requirements for each task. Also, for privacy, several papers tried to define some privacy protection constraints which help to select the best services for each task (e.g., data sensitivity constraints, data usage purpose constraints, data retention time constraints [124]) or restrict the sensitive tasks to be executed in the pre-defined locations (like a private cloud or specific hosts).

### 5.1.3. Audit Mechanisms

These papers report on mechanisms developed to audit the workflow execution and prevent security violations as much as possible. They can be divided into two groups:

a) The ones that only try to detect violations via monitoring. There are three different strategies in terms of the location of the monitoring functionality: a1) Cloud-side Monitoring: These papers rely on a cloud monitoring module in order to detect security violations (e.g. [133], [130], and [47]); a2) Engine-side Monitoring: Violation detection is built into a new module/component of the workflow engine (e.g., [100], [69], [128], and [129]); a3) User-side Monitoring: In this strategy, the user is responsible for monitoring and detecting violations in the workflows based on, for example, a log file (e.g., [126]). Each of these monitoring techniques has its limitations. For example, the first strategy is not fully resistant to some types of attacks. In other words, cloud insiders can misuse the access privileges to undermine the confidentiality, integrity, and availability of the systems [134]. This can be done by performing malicious activities in cloud logs with the aim of destroying attack traces, modifying and deleting log data, diverting the investigation process in other directions so as to hide them, extracting sensitive data, and others [135]. On the other hand, the second method can lead to time and cost overhead for the workflow engine, whereas the last one is often not scalable and can only be applied for small workflows by skilled users.

b) Those that also try to predict, prevent, or react to violations after detection. In other words, these papers try to: b1) *Predict* violations that may happen in the future by extracting knowledge from the security logs and past violations, b2) *React* to them by using recommended actions to eliminate or reduce the effects of already occurred violations in real-time (Full or Partial Compensation), or b3) *Prevent* them from happening by detecting any violation sign.

More details about these methods can be found in section 5.3.

Figure 6 also shows the percentages of each security solution proposed by the papers disaggregated by the type of the workflow. It should be noted that because most of the papers have dealt with more than one objective with more than one solution, Figure 6b shows the number of unique papers in each category.

### 5.2. Workflow Life Cycle and Security

In Figure 7, we show the workflow life cycle that also includes the phase of workflow adaptation. It also visualizes the coverage of the selected papers in the SLR regarding the security and privacy concerns per phase disaggregated by the type of workflow. As shown in the figure, most of the papers focused only on the execution phase for the scientific workflows. In the modeling and IT refinement phases, just a very few research works tried to address the security and privacy requirements in scientific workflows, whereas almost all papers considering business workflows focus on this phase. One of the reasons is that scientists do not distinguish between workflow models and executions. They develop their workflows in a trial-and-error manner and hence the modeling and execution phases are not arranged in a strict sequence [10]. As a result, there is a gap in the literature regarding the modeling or specification of security requirements in the scientific workflows in these phases of the life cycle.

Furthermore, as we can see in the same figure (Figure 7), the majority of the papers for both scientific and business workflows focused on security aspects at the modeling and execution phases thus leaving a huge open research space in the monitoring, analysis, and especially in the adaptation phase of workflow in cloud environments. More specifically, to the best of our knowledge, there is no paper proposing a solution to the adaptation of the currently running instances of business workflows upon security violations. Most of the papers that address the modeling phase (which are mostly also including IT refinement), focused on extending language and modeling tools to specify/capture user security requirements at different levels of abstraction. Many of them define extensions to the widely known BPMN (Business Process Model and Notation) to support the specification of the non-functional requirements such as security. These works used graphical interfaces or textual notations to enrich BPMN diagrams with security properties [136]. Furthermore, some of these research works tried to define initial approaches in order to meet various security goals. For example in [34], firstly, the security goals are described in degrees (e.g., "high" and "low"), and also some actions are defined to meet these different degrees of security. Then, the predefined actions are enforced based on the security goal degree of each task. For example, for the tasks with a high level of confidentiality, the more secure cryptography algorithm, authentication, and access restriction must be applied, whereas for the low-level ones, a less secure cryptographic algorithm is sufficient. Considered from a different perspective, the work [78] models and analyzes the security threats for each security goal and proposes security requirements based on these threats. For instance, they defined certain threats such as the disclosure of information for confidentiality and then proposed some requirements (like data obfuscation) as solutions to prevent these threats. Besides these extensions, some papers tried to consider the cloud features in the modeling phase and model the processes based on these features. These papers tried to use BP obfuscation in order to preserve the privacy of processes in the cloud. Moreover, after the modeling phase, some papers check the models in terms of information flow or control flow via various model checking techniques (discussed in section 5.1 in detail).

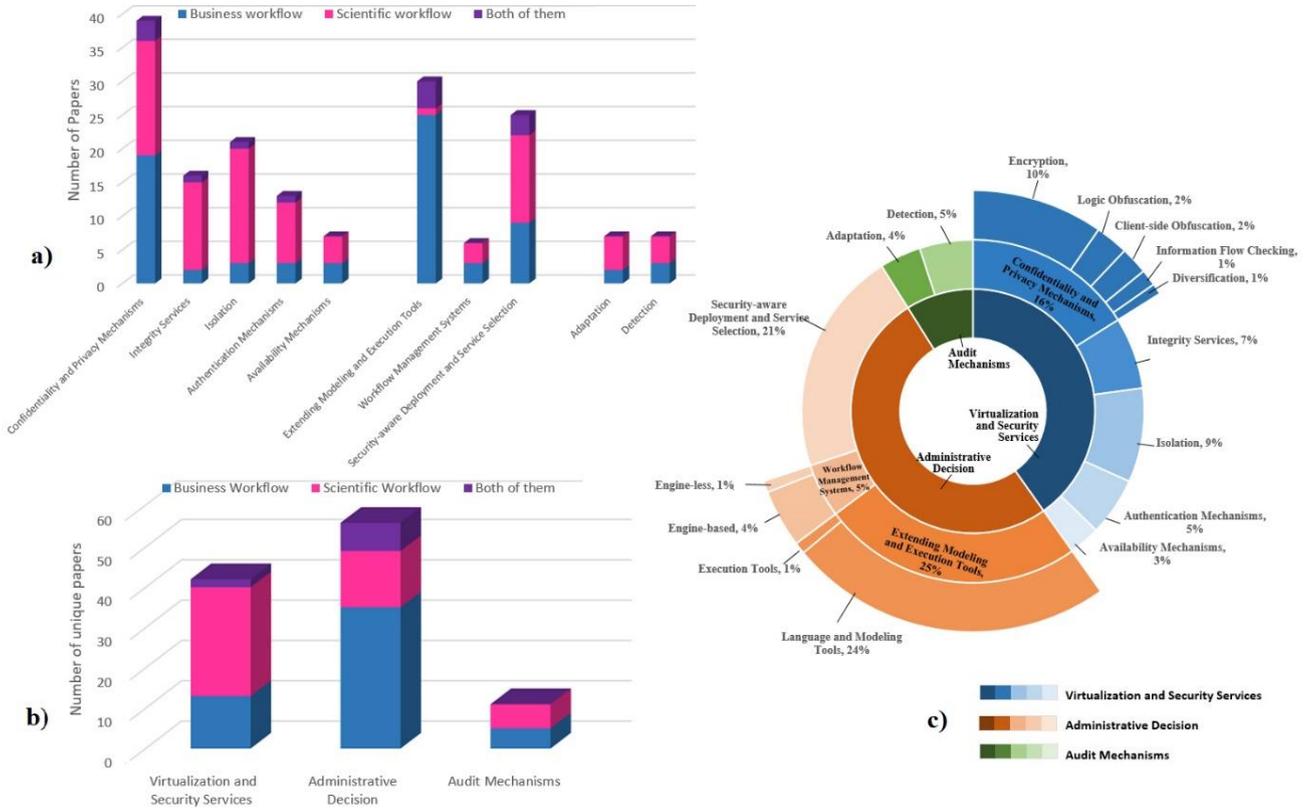

Figure 6: Percentage of Security Solutions proposed by the papers disaggregated by the type of workflow, i.e. business, scientific workflows, or both types. a) Percentage of each subcategory disaggregated by the type of workflow, b) The number of unique papers in each category disaggregated by the type of workflow, c) Percentage of each solution in detail.

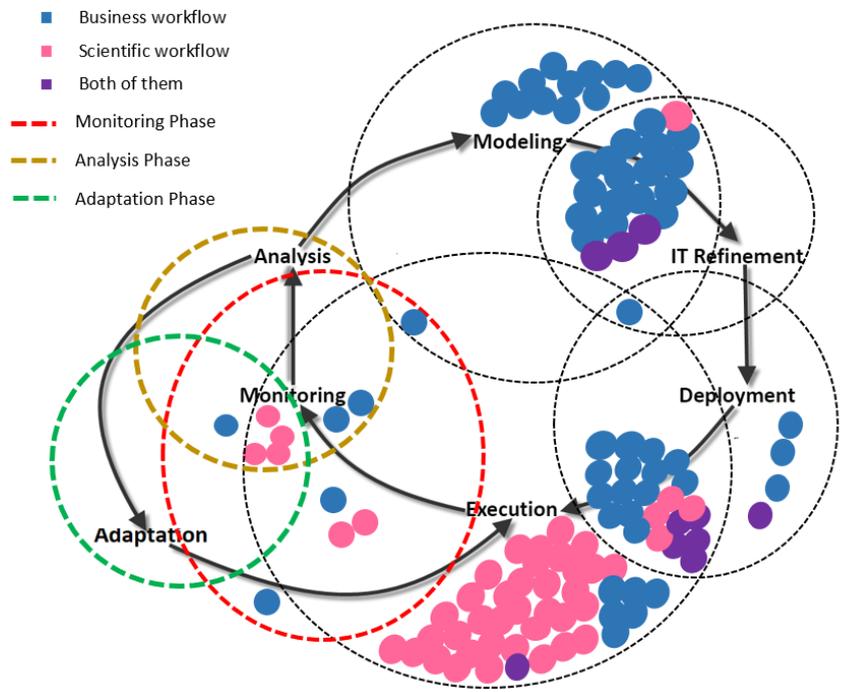

Figure 7: Workflow life cycle coverage by the available security and privacy-related publications disaggregated by the type of workflow.

In the deployment and execution phase, the works propose the selection of the best fit services for each task and apply security services to them based on the user security requirements. At the same time, the aim is to find a balance between these requirements and other user preferences. They often used decision-making criteria including trust, reliability, reputation, or risk assessment techniques in order to find the best fit services.

In the monitoring, analysis, and adaptation phases, the papers mostly used audit mechanisms to monitor the execution of processes, and then based on the obtained information, they aimed at preventing violations or reacting to them. These papers are discussed in more detail in section 5.3

In Figure 8, we present the percentages of the selected papers per workflow life cycle phase.

## 5.3. Monitoring, Analysis and Adaptation

In this section, we briefly discuss the papers that addressed the security concerns in the monitoring, analysis, and adaptation phases. The papers can be categorized based on the type of adaptation they focus on. We identify the following categories, as inspired by [137]: 1) Diagnostics; 2) Predictive Adaptation; 3) Prescriptive Adaptation; 4) Proactive Adaptation. Table 4 shows the summary of the 10 papers, which are the result of our last stage of selection in the research methodology (see Figure 3). The works belonging to each of these groups are summarized in the following subsections.

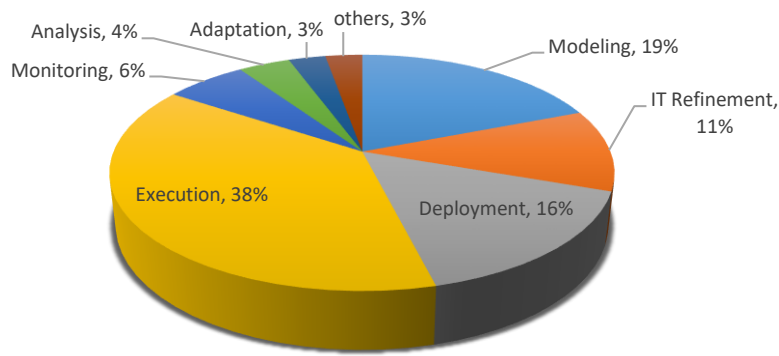

Figure 8: The percentage of selected papers per workflow life cycle phase (total number of papers=120).

Table 4: The summary of the related papers in the monitoring, analysis, and adaptation phases in the cloud.

| Paper | Adaptation Type | | | | Type of Workflow | Trigger (adaptation reason) | Monitoring and Detecting Module |
|---|---|---|---|---|---|---|---|
| | **Diagnostics (Detect)** | **Predictive Adaptation (Predict)** | **Prescriptive Adaptation (React)** | **Proactive adaptation (Prevent)** | | | |
| [126], 2011 | Verifying the Integrity of the computing platform and the correct execution of the outsourced processes by the user. | - | - | | Business workflows | - | The user verifies the log file (User-side Monitoring) |
| [100], 2015 | Input validation (detecting unauthorized input), remote access validation, and data integrity. | - | - | | Scientific workflows | - | Security Analysis Package (SAP) embedded in the Kepler WfMS (Engine-side Monitoring) |
| [127], 2013 | Detecting security issues (information leakage, Man-in-The-Middle (MiTM) attacks, DoS attacks, and resource misuse that can affect confidentiality, integrity, and availability of information) | - | - | | Business workflows | - | Vulnerability Diagnostic Tree (VDT) model by using the audit log (it can be Engine-side or Cloud-side Monitoring) |
| [133], 2020 | - | Minimize the security risk and improve the future composite decisions using this log information | - | | Business workflows | Security issues (availability, integrity, and confidentiality) | Assumption that all of the abnormal behaviors (security attacks) are logged in a broker module (Cloud-side Monitoring) |
| [69], 2016 | Detecting any KPI violations because of the peak-loads period | Find a near-optimal solution based on this information to avoid further violations while considering security requirements. | - | | Business workflows | KPI violations | Engine-side Monitoring |
| [128], 2020 [129], 2021 | Detecting the sub-tasks with low confidence based on the lagged decision mechanism | - | Preserving the intermediate data for re-execution of sub-tasks with low confidence and avoiding the execution of the workflows from the beginning. | | Scientific workflows | The sub-tasks with low confidence | Lagged Decision Mechanism embedded in the workflow monitor module (Engine-side Monitoring) |

| Paper | Adaptation Type | | | | Type of Workflow | Trigger (adaptation reason) | Monitoring and Detecting Module |
|---|---|---|---|---|---|---|---|
| | Diagnostics (Detect) | Predictive Adaptation (Predict) | Prescriptive Adaptation (React) | Proactive adaptation (Prevent) | | | |
| [130], 2020 | - | - | Rescheduling the uncompleted services. | | Scientific workflows | Change of cloud resources availability (like clouds fail and availability of new clouds) | The Cloud Monitor module (Cloud-side Monitoring) |
| [47], 2018 | - | - | Rescheduling the affected tasks | | Scientific workflows | Any kind of malicious behavior in a VM | The Cloud Administrator module |
| [132], 2021 | Detecting failure of tasks | - | Re-executing the failure tasks | - | Scientific workflows | Resource unavailability, an expiry of the deadline | Engine-side Monitoring |
| [131], 2021 | Detecting a high or very high state of future resource load by predicting the resource load status in each VM | - | A live VM migration to transfer the current load of the damaged VM to another VM | Balancing the processing load in resources dynamically | Scientific workflows | Unavailability of VMs | Engine-side Monitoring |

### 5.3.1. Diagnostic

The research works in this group focused only on the detection of the security violations during the monitoring phase of the workflows. The contributions per publication are summarized below.

The authors of [126] provided an architecture that allows users to verify the correctness of the business process executions remotely. For this purpose, they used a mechanism to log sensitive activities of business processes. After the process is completed, the client can request a signed version of the log and check it periodically to verify the correctness of the process execution. This paper did not consider real-time monitoring during execution and automatic adaptation.

In [100], the authors used provenance information for security purposes. They extended the Kepler provenance module and added the Security Analysis Package (SAP) to it in order to analyze provenance information in the security context. They focused on three data-flow oriented security properties: 1) input validation: using a whitelist of acceptable inputs to detect and filter unauthorized input, 2) remote access validation: implementing an internal firewall that contains the valid URLs and IP addresses, and 3) data integrity: comparing on-hash and post-hash of data to check data integrity. However, the provenance information was only used for detecting a few security violations and defining a follow-up action for the future scientific workflow executions in an attempt to prevent attacks.

In [127] the authors used cloud-wide auditing to uncover security issues. They defined Vulnerability Diagnostic Trees (VDTs) to formally manifest vulnerability patterns across several audit trails. This method can be used for implementing automated detection mechanisms that take various audit trails as input and pinpoints threats using their type and location. However, they did not consider the different requirements of users and services during their diagnostics. Furthermore, this method is not scalable as investigating all possible attacks using this method for all services can bring unnecessary time and computing overhead to the system reducing the scope of the work down to only a few possible attacks and violations in the service composition. Clearly, this paper only detected some violations based on the audit log and could not prevent or react to them.

### 5.3.2. Predictive Adaptation

The research works in this category attempted to extract knowledge from the security logs to predict the future states, outcomes, or properties of a process instance or group of process instances and then prevent or reduce the further violations based on these predictions and improve future decisions. In other words, these methods try to use dynamic scheduling based on real-time information to prevent further violations in the new instances. Therefore, they cannot prevent attacks in already running instances.

[133] used the providers' security log (which registers cloud anomalies such as cybersecurity attacks) to quantify the Cloud Security Risk (CSR) regarding the user's business process and then formulated the web service composition problem as a bi-objective optimization problem with service cost and multi-cloud risk viewpoints. We categorize this work in the Diagnostics group as it uses log information to improve the process design by minimizing the security risk and hence diagnoses potential issues with a cloud provider. However, this work may also be considered as a prescriptive approach, as it prescribes the design of a service composition no adaptation taking place in this work.

[69] presented an adaptive and context-aware decision system that can predict peak-load times of business processes where the KPI thresholds are mostly surpassed/violated. The authors use a decision tree technique which is applied to the business process execution log to extract this knowledge. Then, the penalty-based genetic algorithm is applied to find a near-optimal solution while considering the required level of security for each task. However, like the previous paper, this work also cannot adapt the currently running process instances.

### 5.3.3. Prescriptive Adaptation

The papers discussed here aim at detecting violations and automatically recommending actions to prevent or reduce the effects of already occurred violations in real-time. In other words, these works focus on the adaptation of already running instances in order to fully or partially compensate for the effect of the occurred violations by taking proper actions.

The authors of [128] proposed an intrusion tolerant scientific workflow system. They used different techniques to secure the workflow execution: 1) Executing the same sub-task in parallel in multiple heterogeneous Virtual Machines (VMs) to enhance reliability; 2) Proposing the dynamic task scheduling strategy based on resource circulation to cut off the attack chain; 3) Designing the temporary workflow intermediate data backup strategy. technique that can avoid executing the compromised workflows from the beginning. Furthermore, this paper only focused on detecting the integrity violations during the execution phase of scientific workflows and could not This preserved the intermediate data which can be used for the re-execution of the workflow sub-tasks with low confidence; the confidence of tasks is assessed by the so-called lagged decision mechanism as the assessment is provided after the tasks have been executed. In [129], the same authors tried to solve a limitation of the previous work by scheduling the replicas of sub-tasks so that the attacker cannot destroy the whole workflow just by compromising one VM. This method can be considered as an adaptable discover the other types of security violations.

The research presented in [130] introduced a dynamic rescheduling method to handle the changes in cloud resources availability like the failure of the existing resources or the availability of new resources. In other words, if a changed status is detected, the cost model dynamically calculates the cost of deploying uncompleted tasks onto the currently available cloud resources and reschedules them to handle run-time failures. Note that resource availability during the execution of the scientific workflow is the only security factor considered by this work.

An attack response approach to reduce the security threat in scientific workflows is introduced in [47] in which the security threat is calculated by considering one task to be malicious and then estimating the number of affected tasks (using a simple variant of the decision tree). Then after detecting any kind of malicious behavior in a VM by the cloud administrator, the attack response module tries to reschedule the high-risk tasks (including the running tasks and future tasks). The paper only addressed the integrity and availability of the data during the execution of scientific workflows and no other security requirements are considered.

### 5.3.4. Proactive Adaptation

These methods try to predict the violations before they occur and then adapt the already running instances in order to prevent these violations from happening.

The work presented in [131] proposed a combination of two adaptation approaches during the scheduling of the scientific workflows in order to tolerate faults (i.e., the unavailability of VMs): 1) Proactive Adaptation: Applying a prediction model to proactively control resource load fluctuation. In better words, this model can increase the failure prediction accuracy before fault/failure occurrence. 2) Prescriptive Adaptation: Applying a reactive fault tolerance technique for when a processor fails and the scheduler must allocate a new VM to execute the workflow tasks. However, this paper only focused on the VM faults and did not address other security issues.

## 5.4. Workflow Management System

In this section, we provide details about the contributions that mention the development of new engine-workflow management systems or extend the already existing ones to handle the security requirements.

[27] proposed a framework of, what they call, a "mimic cloud workflow execution system" with three strategies: heterogeneity (diversification of physical servers, hypervisors, and operating systems), redundancy (Lagged Decision Mechanism), and dynamics (switching workflow execution environment). However, this system only covered the execution and monitoring phases of the workflow life cycle and cannot carry out adaptation of the process instances.

[52] developed a secure big data workflow management which they called SecDATAVIEW, based on DATAVIEW [138]. This system leverages the hardware-assisted trusted execution environments (TEEs) such as Intel Software Guard eXtensions (SGX) and AMD Secure Encrypted Virtualization (SEV) to protect the execution of big data workflows and the data used by them. They also proposed a secure architecture and the WCPAC (Workflow Code Provisioning and Communication) protocol for securing the execution of workflow tasks in remote worker nodes. However, this system is still vulnerable to some attacks including network traffic-analysis, denial-of-service, side-channel attacks, and fault injections. Furthermore, it only protects workflows from possible attacks during execution, and if an attack occurs, it terminates the workflow execution. In other words, there is no adaptation module in this system.

[100] extended the Kepler provenance module and added the Security Analysis Package (SAP) to it in order to analyze provenance information in the security context using three security properties (i.e., input validation, remote access validation, and data integrity). As mentioned earlier, this module can only detect a few violations and cannot adapt running workflows to react to security violations.

[34] proposed an integrated environment named BPA-Sec4Cloud, which aims to provide a holistic and integrated cloud-based solution to address the automation of security-aware business processes from its modeling to its deployment. For instance, as already discussed in Section 5.2, they used a BPMN-based Editor with security abstractions and service modeling capabilities support. However, the monitoring, analysis, and adaptation phases are not supported by the system.

[76] presented a cloud workflow engine based on the extended jBPM4 (Java Business Process Management) that can support privacy protection between different tenant workflow instances in the cloud workflow systems. They defined three levels of isolation: 1) data isolation: protecting the private data produced in the execution process of tenants' workflow instances; 2) performance isolation: protecting the process instance information belonging to different tenants at run-time; 3) execution isolation to meet the different performance requirements for various tenants. However, they cannot detect the security violations in workflows and like others, did not consider the adaptation phase.

These WfMSs are summarized in Table 5. As the table shows, there is no WfMS that can adapt running workflows instances in order to prevent security violations or react to them. Furthermore, none of the existing WfMSs are able to handle all of the security concerns during the whole workflow life cycle.

Table 5: Different WfMSs regarding security concerns in the cloud.

| WfMS | Type of supported Workflow | Supported Representation Model | Extension | Execution Environment | Covered Security Objectives |
|---|---|---|---|---|---|
| [27], 2018 | Scientific | DAG | - | Cloud-based | • Data Integrity<br>• Data Confidentiality |
| SecDATAVIEW [52], 2019 | Scientific | DAG | DATAVIEW | Execute the kernel of WfMS (the components that process confidential data) inside SGX enclaves (other components are executed on the trusted premises such as private cloud computing platforms or the user side premise). | • Data/Task Integrity<br>• Data/Logic Confidentiality |
| [100], 2015 | Scientific | DCG | Kepler | Flexibly arranged to run locally or on a cloud platform | • Data Integrity |
| BPA-Sec4Cloud [34], 2016 | Business | BPMN | - | Cloud-based | • Data Confidentiality<br>• Data Integrity<br>• Authentication |
| [76], 2019 | Business | BPMN | jBPM4 | Cloud-based | • Data/Logic Confidentiality |

### 5.5. Findings and Future Work

In this section, we briefly summarize our main findings and discuss some of the challenges and open issues in different phases of the workflow life cycle for further improvement.

Our first finding is related to the need for automation in the modeling and translation of the security requirements. For this purpose, we identify a need for a standard workflow modeling notation that can cover all security requirements and then clear and standard mapping rules in order to define specific solutions for the intended security goals. This automation can facilitate the understanding and use of the system even by non-security specialists for the purposes of conflict analysis, reuse, and validation of the model. However, each of the existing security-aware modeling languages, like the existing BPMN extensions, proposes its model and notation which creates a barrier to their adoption. Besides, despite existing formal specifications of security goals/properties and available security patterns, there is a lack of automated model checking for satisfying the workflow's security requirements and quantifying the information leakage and control flow risks between tasks. Beyond that, because of the growing interest in adopting the cloud for workflow execution, we need to pay attention to the cloud characteristics during the workflow modeling. It can produce workflows that are more adaptable in terms of not only security and privacy concerns but also other parameters like efficiency and cost. Up to now, there are only a few research works that try to consider the characteristics of Clouds by using some mechanisms like Client-side Obfuscation or BP Obfuscation before sending workflows to the cloud. Therefore, we can conclude that there is a lack of research work that accounts for the cloud features during workflow modeling towards specifying all relevant information (using modeling tools and languages) for cloud outsourcing of processes.

In the deployment and execution phase, since most security challenges in cloud infrastructure are due to Virtualization Technology (VT), it is important to consider the relation between virtualization technology and security and select the right VT. This can provide QoS to the end-users with lower prices and also a cost-effective solution and efficient resource utilization to the Cloud Service Providers (CSP). So, there is a need of proposing a security-aware scheduling method for selecting the proper VT (like Virtual machines (VM), Virtual Containers (VCs), Containers within VM, Lightweight VM, or Unikernel [139]) at the task-level and also at the workflow-level based on the workflow's characteristics and users' requirements.

As our main finding in this literature review, there is a big gap in research regarding the monitoring, analysis, and adaptation phases. In the monitoring phase, there is no reliable and scalable strategy that can detect all kinds of possible attacks during the workflow execution. Besides, in the analysis and adaptation phases, there is a huge gap in preventing and reacting to security and privacy violations. In particular, for the business workflows, there is almost no paper dealing with adapting workflows regarding these violations. Also, in the scientific context, just a few research works tried to adapt workflow considering only one or two types of violations. Therefore, it is important to develop techniques for both scientific and business workflows that can detect, prevent and react to security violations and compensate for part or all of the damage.

Finally, to the best of our knowledge and based on this review of the literature, there is no cloud-based WfMS that can handle the security and privacy concerns during the whole workflow life cycle in both business and scientific workflows. Especially regarding the adaptation phase, there is a significant gap in existing WfMSs. In the modeling phase, the WfMS should be able to extend a model that can capture the fundamental security principles of CIANA in a clear and precise way. Next, in the execution phase, the system needs some mechanisms to make the right choices of the wide range of offers in the cloud environments based on the specified demands. Moreover, in the monitoring phase, the system must monitor the instances and the execution environment to detect security violations (based on the Service-Level Agreement (SLA) or Key Performance Indicator (KPI)), unexpected behavior, failure of a task or instance, unavailability of a service or resource and so on. After this detection, it is important to provide workflow flexibility to react to such violations appropriately and also prevent them from propagating in the whole workflow execution. In other words, we need a WfMS with a detection and adaptation module that can handle any possible violations. Also, such a cloud-based system should provide an isolated execution and monitoring environment for each user to configure and run its workflow and meet various functional and non-functional requirements of the different users. Moreover, in terms of scalability, there is a gap in research for a security-aware distributed or even engine-less WfMS in the cloud environments that can be more compatible with the cloud characteristic.

## 6. Conclusions

This paper presents a comprehensive overview of the security and privacy properties in cloud-based business and scientific workflows. We used a two-step approach with which we aimed at identifying the main challenges and potential research directions in this area of research. The first step followed the SMR protocol in which we classified the available literature based on the proposed security solutions and their target phases in the workflow life cycle. To do so, we devised three categories covering various aspects of security and privacy concerns in workflows: Virtualization and Security Services, Administrative Decision, and Audit Mechanisms. Our findings show that most of the available relevant literature focuses predominantly on the modeling phase and less so on the execution phase. We noticed that there is a lack of sufficient attention to the monitoring, analysis, and adaptation phases for both business and scientific workflows, while the adaptation phase is in fact neglected up till now. As an additional result of this step we identified a total of 120 publications that we investigated in the second step of our study.

In the second step we followed the SLR protocol which we used on the set of publications identified by the SMR and the corresponding gaps in the state of the art related to security concerns in the monitoring, analysis, and adaptation phases of workflows. We compared the works using the criteria adaptation type, type of considered workflow, adaptation reason, and monitoring and detection module/mechanism. Based on that, we conclude that there is a gap in the state of current research on reliable and scalable approaches that can detect, prevent and react to security violations and compensate for part or all of the damage for both scientific and business workflows that are cloud-based. Moreover, we investigated the existing WfMSs with respect to the type of supported workflow, the supported representation model, the covered phases of the workflow life cycle, and the covered security objectives. Based on our findings, there is no comprehensive workflow management system in the cloud environments that can handle the security and privacy concerns during the whole workflow life cycle neither in business nor in scientific workflow research. The conclusions of our survey clearly identify a huge potential for future research in approaches and WfMSs that address the security concerns during the whole life cycle of cloud-based workflows and in particular in the monitoring, analysis, and adaptation phases.